\documentclass[twoside]{article}
\usepackage{fleqn,espcrc2,epsf}
\usepackage{graphicx}
\usepackage{epsfig}
\usepackage[figuresright]{rotating}

\newcommand{\ba}[1]{\begin{eqnarray} \label{(#1)}}
\newcommand{\ea}{\end{eqnarray}}

\newcommand{\AmS}{{\protect\the\textfont2
  A\kern-.1667em\lower.5ex\hbox{M}\kern-.125emS}}

\def \znbb {$0\nu\beta\beta$}

\def\be{\begin{equation}}
\def\ee{\end{equation}}
\def\bea{\begin{eqnarray}}
\def\eea{\end{eqnarray}}

\title{Search for Dark Matter by GENIUS-TF and GENIUS}

\author{H.V. Klapdor-Kleingrothaus
\address{Max-Planck-Institut f\"ur Kernphysik, P.O. Box 10 39 80,\\
 D-69029 Heidelberg, GERMANY}\thanks{Spokesman of HEIDELBERG-MOSCOW and GENIUS Collaborations
	{\it E-mail}: klapdor@gustav.mpi-hd.mpg, 
 {\it Home-page}: http://www.mpi-hd.mpg.de.non$\_$acc/}
}

\begin{document}

\begin{abstract}
	The new project 
	GENIUS will cover a wide range of the parameter 
	 space of predictions of SUSY for neutralinos as cold dark matter. 
	Together with DAMA it will be the only experiment 
	which can probe the seasonal modulation signal. 
	Concerning hot dark matter GENIUS will be able 
	to fix the (effective) neutrino mass with high accuracy.
	 A GENIUS Test Facility has just been funded and will 
	 come into operation by end of 2002.
\end{abstract}

\maketitle
\section{GENERAL SITUATION}

	Dark matter is at present one of the 
	most exciting fields of particle physics and cosmology. 
	Recent investigations of the CMB (MAXIMA, BOOMERANG, DASI) 
	together with large scale structure results 
	fix the non-baryonic dark matter contribution 
	to the mass of the Universe to $\sim$ 25$\%$, 
	of which still up to 38$\%$ could be hot (neutrinos). 
	With the early nucleosynthesis constraint of $\Omega$= 0.04, 
	the overwhelming contribution would consist of dark energy.

	Terrestrial direct dark matter search experiments 
	just now are {\it starting} to enter into the range 
	of sensitivity required, it we consider the neutralinos 
	to be the candidate for cold dark matter. 
	The situation is similar in the hot Dark Matter Search. 
	Here double beta decay is the most sensitive means 
	to look for an absolute neutrino mass scale.

	Direct search for WIMPs can be done

	(a) by looking for the recoil nuclei 
	in WIMP- nucleus elastic scattering. 
	The signal could be ionisation, phonons or light produced 
	by the recoiling nucleus. 
	The typical recoil energy is a few 100\,eV/GeV WIMP mass.

	(b) by looking for the modulation of the WIMP  
	signal resulting from the seasonal variation 
	of the earth's velocity against the WIMP 'wind'.

	The expectation for neutralino elastic scattering 
	cross sections and masses have been extensively 
	analysed in many variants of SUSY models.


\begin{figure}[h]
\begin{center}
\epsfig{file=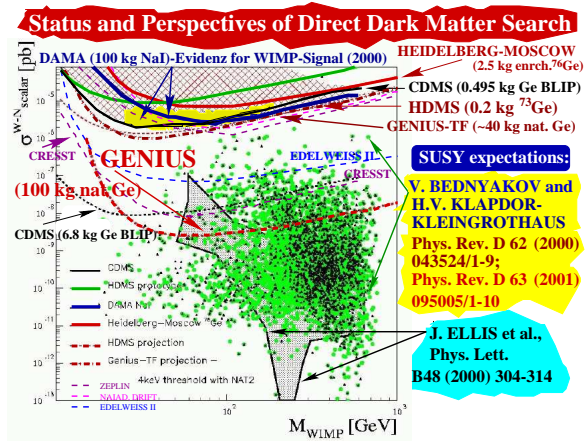,scale=0.3}
\end{center}

\vspace{-0.5cm}
\caption[]{
       WIMP-nucleon cross section limits in pb for scalar interactions as 
       function of the WIMP mass in GeV. 
       Shown are contour lines of present experimental limits (solid lines) 
       and of projected experiments (dashed lines). 
       Also shown is the region of evidence published by DAMA. 
       The theoretical expectations are shown 
	for the MSSM by two scatter plots, 
	- for accelerating and for non-accelerating Universe (from  
\cite{BedKK00_01}) 
	and for MSUGRA by the grey region (from  
\cite{EllOliv-DM00}). 
	{\em Only}~ GENIUS will be able to probe the shown range 
       also by the signature from seasonal modulations.
\label{fig:Bedn-Wp2000}}
\end{figure}


	Fig. 
\ref{fig:Bedn-Wp2000}
	represents the present situation. 
	The SUSY predictions are from the MSUGRA model 
\cite{EllOliv-DM00}
	and the MSSM with relaxed unification conditions 
\cite{BedKK00_01}. 
	Fig. 
\ref{fig:Ell-New-01-11}
	shows the result of a study 'at Post-LEP Benchmark points'
	based again on the MSUGRA 
\cite{Ell-New-01-11}. 
	Present experiments only just touch the border 
	of the area predicted by the MSSM. 
	The experimental DAMA evidence for dark matter 
	lies in an area, in which MSUGRA models do not expect dark matter. 
	They would require Beyond-GUT physics in this frame  
\cite{Arn-priv01}.

	Summarizing the present experimental status, 
	present experiments and also future projects (see  
\cite{KK-LP01})
	can be categorized in two classes:

	1. Sensitivity (or sensitivity goal) 'just for' confirmation of DAMA.

	2. Sensitivity to enter deeply into the range of SUSY predictions.

	Only very few experiments may become candidates 
	for category 2 in a foreseeable future (see Figs.
\ref{fig:Bedn-Wp2000},\ref{fig:Ell-New-01-11},
	and as far as at present visible, of those 
	only GENIUS will have the chance to search for modulation, 
	i.e. to check, like DAMA, positive evidence for a dark matter signal.
	
	The cryogenic experiments CDMS, CRESST and Edelweiss 
	operate at present with 600 g of detectors and less, 
	after a decade of development. 
	Expansion to several tens of kg or better 100\,kg 
	as required for modulation search, is still a {\it very} large step.
	CDMS has collected only 10.6\,kg\,d of data over this time 
	(in 1999, since then not working), Edelweiss only 4.53\,kg\,d.

	The superheated droplet detectors PICASSO/SIMPLE 
	are working at present on a scale of 15\, and 50\,g detectors, 
	respectively.

	ZEPLIN and the very ambiguous and nice project DRIFT 
	are in the stage of prototype construction.

	The situation is different for GENIUS, 
	which is based on conventional techniques.


\section{GENIUS AND COLD DARK MATTER SEARCH}

\begin{figure}[htb]

\vspace{-0.5cm}
\centering{
\includegraphics*[width=75mm, height=55mm]{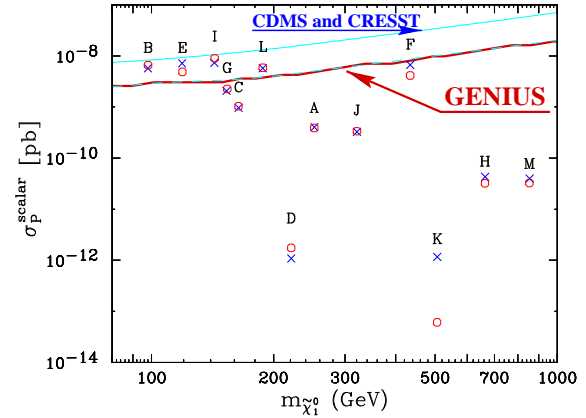}} 

\vspace{-0.5cm}
\caption{WIMP- proton elastic scattering cross sections 
	according to various MSUGRA models (see text). From 
\cite{Ell-New-01-11}.
\label{fig:Ell-New-01-11}}
\end{figure}

	If classifying the SUGRA models 
	into more g$_\mu$-2-friendly (I,L,B,G,C,J) and less 
	g$_\mu$-2-friendly models, according to 
\cite{Ell-New-01-11},
	the former ones have good prospects 
	to be detectable by LHC and/or a 1\,TeV collider. 
	GENIUS could check not only the larger part of these ones, 
	but in addition two of the less 
	g$_\mu$-2-friendly models (E and F), 
	which will be difficult to be probed by future colliders (see Fig. 
\ref{fig:Ell-New-01-11}).

	This demonstrates nicely the complementarity 
	of collider and underground research.

	GENIUS would already in a first step, with 100\,kg of 
		{\it natural} Ge detectors, cover a significant part of the 
		SUSY parameter space for prediction of neutralinos 
		as cold dark matter 
(Figs.~\ref{fig:Bedn-Wp2000},\ref{fig:Ell-New-01-11}). 
	For this purpose the background in the energy range 
		$< 100$\,keV has to be reduced to 
		$10^{-2}$ (events/\,kg\,y\,keV). 
	At this level solar neutrinos as source of background 
	are still negligible. 
	Of particular importance is to shield the detectors 
	during production (and transport) to keep the background 
	from spallation by cosmic rays sufficiently low 
	(for details see 
\cite{KK-NOON00}.

	     The sensitivity of GENIUS for Dark Matter corresponds to 
	     that obtainable with a 1\,km$^3$ AMANDA detector for 
	     {\it indirect} detection (neutrinos from annihilation 
	     of neutralinos captured at the Sun) (see  
\cite{Eds99}). 
	Interestingly both experiments would probe different neutralino 
	compositions: GENIUS mainly gaugino-dominated neutralinos, 
	AMANDA mainly neutralinos with comparable gaugino and 
	Higgsino components (see Fig.\,38 in  
\cite{Eds99}).

\vspace{-0.3cm}
\section{GENIUS AND HOT DARK MATTER SEARCH}

	Neutrinos could still play an important role 
	as hot dark matter in the Universe. 
	Recent results from CMB and LSS measurements still allow 
	for a sum of neutrino masses of $\sum m_i <$ 4.4.\,eV.
	Under the assumption that neutrinos are degenerate in mass, 
	the common mass eigenvalue  
	is, from \znbb~ decay, $<$1.4\,eV\,(90$\%$ c.l.) 
	if the LMA MSW solution is realized in nature 
\cite{KKPS}.
	New approaches and considerably enlarged experiments, 
	will be required in future to fix the neutrino 
	mass with higher accuracy. GENIUS is probably the most 
	straightforward and promising of them. As discussed 
\cite{KK-Erice01,KK-Bey97,GEN-prop,KK-Neutr98,KK-WEIN98,KK-NOW00,KK-J-PhysG98,KK-InJModPh98,KK-SprTracts00,KK60Y}
	it could probe with 1\,ton of enriched $^{76}{Ge}$ 
	a scale of the effective neutrino mass down to 0.02\,eV. 
	This means that with 100\,kg of enriched $^{76}{Ge}$, 
	a scale down to 0.04\,eV could be investigated.
	


\vspace{-0.3cm}
\section{GENIUS-TF}

	As a first step of GENIUS, a small test facility, GENIUS-TF, 
	is at present under installation in the Gran Sasso 
	Underground Laboratory 
\cite{GENIUS-TF}.
	With about 40 kg of natural Ge detectors operated 
	in liquid nitrogen, GENIUS-TF could test the DAMA seasonal 
	modulation signature for dark matter. 
	Up to summer 2001, already six 2.5 kg Germanium detectors with 
	an extreme low-level threshold of $\sim$500 eV have been produced.
	


\vspace{-0.3cm}
\section{CONCLUSION}

	Terrestrial dark matter experiments searching 
	for cold dark matter only now are starting to enter 
	into the range of SUSY predictions. GENIUS and 
	GENIUS-TF could play a decisive role here, 
	as also in fixing the contribution of neutrinos to hot dark matter.


\end{document}